\documentclass[twocolumn]{article}

\usepackage{arxiv}

\usepackage[utf8]{inputenc} % allow utf-8 input
\usepackage[T1]{fontenc}    % use 8-bit T1 fonts
\usepackage{framed,multirow}
\usepackage{amssymb}
\usepackage{latexsym}
\usepackage{url}
\usepackage{xcolor}
\usepackage{soul,color}
\usepackage{svg}
\usepackage{pdflscape} 
\usepackage[export]{adjustbox}
\usepackage{hyperref}
\definecolor{newcolor}{rgb}{.8,.349,.1}
\usepackage{booktabs, multirow, tabularx}
\usepackage{subcaption}

\title{Rewiring Development in Brain Segmentation: Leveraging Adult Brain Priors for Enhancing Infant MRI Segmentation}

\author{
Alemu Sisay Nigru  $^\diamond$\\
Department of Information Engineering\\
University of Brescia (Italy)
\And
Michele Svanera$^\diamond$\thanks{Corresponding author: Michele.Svanera at glasgow.ac.uk - $^\diamond$ Shared authorship.}\\
School of Psychology \& Neuroscience\\
University of Glasgow (UK)
\And
Austin Dibble\\
School of Psychology \& Neuroscience\\
University of Glasgow (UK)
\And
Connor Dalby\\
School of Psychology \& Neuroscience\\
University of Glasgow (UK)
\And
Mattia Savardi\\
Department of Medical and Surgical Specialties, \\
Radiological Sciences, and Public Health\\
University of Brescia (Italy)
\And
Sergio Benini\\
Department of Information Engineering\\
University of Brescia (Italy)
}

\begin{document}
\maketitle
\vspace{-10cm}

%%%%%%%%%%%%%%%%%%%%%%%%%%%%%%%%%%%%%%%%%%%%%%%%%%%%%%%%%%%%%%%%%%%%%%%%%%%%%%%%%%%%%%%%%%%%%%%%%%%%%%%%%%%%%%%%%%%%%%%%%%%%%%
%{\onecolumn
\twocolumn[
\begin{abstract}

Accurate segmentation of infant brain MRI is critical for studying early neurodevelopment and diagnosing neurological disorders.
Yet, it remains a fundamental challenge due to continuously evolving anatomy of the subjects, motion artifacts, and the scarcity of high-quality labeled data. 
In this work, we present LODi, a novel framework that utilizes prior knowledge from an adult brain MRI segmentation model to enhance the segmentation performance of infant scans. 
Given the abundance of publicly available adult brain MRI data, we pre-train a segmentation model on a large adult dataset as a starting point. 
Through transfer learning and domain adaptation strategies, we progressively adapt the model to the 0-2 year-old population, enabling it to account for the anatomical and imaging variability typical of infant scans.
The adaptation of the adult model is carried out using weakly supervised learning on infant brain scans, leveraging silver-standard ground truth labels obtained with FreeSurfer. 
By introducing a novel training strategy that integrates hierarchical feature refinement and multi-level consistency constraints, our method enables fast, accurate, age-adaptive segmentation, while mitigating scanner and site-specific biases. 
Extensive experiments on both internal and external datasets demonstrate the superiority of our approach over traditional supervised learning and domain-specific models. 
Our findings highlight the advantage of leveraging adult brain priors as a foundation for age-flexible neuroimaging analysis, paving the way for more reliable and generalizable brain MRI segmentation across the lifespan.

\keywords{3D segmentation \and brain MRI \and level-of-detail architecture \and multi-site learning \and infant brain}
\end{abstract}
\bigskip
]

%\linenumbers
%%%%%%%%%%%%%%%%%%%%%%%%%%%%%%%%%%%%%%%%%%%%%%%%%%%%%%%%%%%%%%%
%%%%%%%%%%%%%%%%%%%%%%%%%%%%%%%%%%%%%%%%%%%%%%%%%%%%%%%%%%%%%%%
\section{Introduction}
\label{sec:introduction}

%importance of first year
The first year of human life represents a pivotal phase in postnatal brain development, characterized by rapid tissue growth and the emergence of diverse cognitive and motor functions \cite{li2013mapping, li2014mapping}.  
The increasing availability of non-invasive multimodal magnetic resonance imaging (MRI) of infant brains has opened remarkable opportunities for mapping early neurodevelopmental trajectories with high precision.  
Large-scale initiatives such as the Baby Connectome Project (BCP) \cite{howell2019unc} and the Developing Human Connectome Project (dHCP) \cite{makropoulos2018developing} have significantly enhanced our understanding of normative brain growth across key developmental stages, spanning from birth to 5 years in the former and from 20 weeks of gestation to 44 weeks post-menstrual age in the latter.  
These datasets offer invaluable insights into both typical brain maturation and atypical trajectories associated with neurodevelopmental disorders, including autism \cite{li2018early}.  

%importance of segmentation
Accurate segmentation of infant brain MR images into distinct brain areas such as white matter (WM), gray matter (GM), and cerebrospinal fluid (CSF), is essential for studying both typical and atypical early brain development \cite{wang2014segmentation, wang2015links}.  
Segmentation plays a foundational role in downstream processing tasks, including image registration \cite{hu2017learning} and atlas construction \cite{shi2010construction, shi2014neonatal}.  
However, infant brain MRI presents unique challenges due to the dynamic nature of early neurodevelopment.  
In the infantile phase ($\leq$ 5 months), T1-weighted (T1w) images exhibit higher signal intensity in GM than in WM.  
During the isointense phase (6-9 months), myelination and maturation lead to increased WM intensity in T1w images, reducing the contrast between GM and WM.
This phase represents the most challenging period for segmentation due to substantial intensity overlap between GM and WM voxels, the increased motion artifacts typical in infant scans, the nonlinear nature of early brain development \cite{paus2001maturation}, and significant partial volume effects.  
Beyond 9 months, the brain begins to exhibit an early adult-like instensity distribution, where WM appears brighter than GM in T1w images.  

%lack of tools
Existing computational tools designed for adult brain MRI, such as SPM \cite{penny2011statistical}, FSL \cite{jenkinson2012fsl}, BrainSuite \cite{shattuck2002brainsuite}, CIVET \cite{ad2006civet}, FreeSurfer \cite{fischlFreeSurfer2012a}, and the HCP pipeline \cite{glasser2013minimal}, often perform suboptimally when applied to infant brain MRI \cite{li2019computational}.  
Currently, only a few dedicated pipelines exist for infant brain MRI processing: among them, the dHCP minimal processing pipeline \cite{makropoulos2018developing} and Infant FreeSurfer \cite{zollei2020infant}.
Prior to the introduction of segmentation methods developed for the \href{http://iseg2017.web.unc.edu/}{iSeg-2017} \cite{wang2019benchmark} and \href{https://iseg2019.web.unc.edu/}{iSeg-2019} \cite{sun2021multi-infant} challenges, only \href{http://www.ibeat.cloud}{iBEAT V2.0 Cloud} \cite{wang2023ibeat} was capable of partially handling the isointense phase.  
However, this tool suffers from high computational demands and a limited number of segmented structures.  

%challenges
Accurately segmenting infant brain tissues presents considerable challenges, primarily due to three key factors.
First, the availability of annotated infant clinical data is limited: \textit{manual annotation} is particularly difficult in early infancy, especially for neonates younger than three months, due to low tissue contrast that complicates the delineation of anatomical structures.
Second, improving \textit{model generalization and portability} across different imaging sites requires training on a diverse set of infant MR scans acquired from multiple institutions.
However, leveraging infant images acquired using varying magnetic field strengths, head coils, and imaging protocols introduces a significant challenge known as the \textit{scanner effect} \cite{svanera2024fighting}.
Deep learning (DL) models often struggle to generalize to unseen imaging sites, exhibiting a notable drop in performance due to inconsistencies in acquisition parameters.
Last, while large-scale open datasets are widely available for adult brain imaging, publicly accessible infant datasets remain scarce, further constraining model development and validation.

%%%%%%%%%% SB UP TO HERE %%%%%%%%%%%%%%%%%

\subsection{Main contributions}

This work presents LODi (Level-Of-Detail-infant) a novel approach which leverages adult brain data as a prior for accurate infant brain MRI segmentation. 
Unlike traditional age-specific models that require extensive labeled datasets for each developmental stage, LODi transfers anatomical knowledge from adults to infants through a carefully designed deep learning pipeline. 
This specifically-designed architecture integrates adult brain anatomical priors learned from a large-scale adult MRI dataset of 27k scans  \cite{svanera2024fighting},  to improve segmentation of infant data.  
Through a multi-stage training procedure, we first pre-train a hierarchical segmentation model on adult brain scans to establish a robust feature representation; second, we transfer the model using weakly supervised learning on infant MRI, guided by silver-standard ground truth annotations generated with Infant FreeSurfer \cite{zollei2020infant}. 
While the adult brain prior captures fundamental structural patterns that remain consistent across the human lifespan \cite{bontempi2020cerebrum, svanera2024fighting}, the transfer stage allows flexibility to adapt to age-specific anatomical variations and intensity distributions. 
This strategy significantly improves segmentation accuracy, especially in the context of scarce and heterogeneous infant data, where deep learning methods struggle to generalise, due to data limitations.

Extensive validation across multiple publicly available test sets, confirms the superior accuracy, robustness, and generalization capability of LODi compared to conventional age-specific and site-dependent models.
This developmental-aware approach to brain MRI segmentation, where knowledge from adult MRI prior is adapted to a different target age group, has potentially significant implications for longitudinal brain studies, clinical diagnostics, and developmental neuroscience.
%, while keeping the computational impact low.
To maximise research reproducibility and state-of-the-art comparisons, we adopt for testing the MICCAI anatomical structure labels proposed in \cite{mendrikMRBrainSChallengeOnline2015a}, as they offer a standardized, expert-annotated ground truth widely adopted in the literature, ensuring consistency,  and reliable benchmarking.

To ensure full reproducibility and facilitate fair comparisons, we make publicly available our code, a dedicated \href{https://rocknroll87q.github.io/}{project website}, and a containerized environment for ease of deployment. Additionally, we release the complete list of volumes used for training, validation, and testing. This allows other researchers to replicate our experiments and benchmark their methods under exactly the same conditions.

%%%%%%%%%%%%%%%%%%%%%%%%%%%%%%%%%%%%%%%%%%%%%%%%%%%%%%%%%%%%%%%
\section{Related work}
\label{sec:soa}

Prior to the introduction of DL-based methods, the dHCP minimal processing pipeline \cite{makropoulos2018developing} and Infant FreeSurfer \cite{zollei2020infant} were the reference tools in the field.
However, they are limited to either the infantile phase or the early adult-like phase, lacking comprehensive coverage of the entire first-year development.
Therefore, in the last decade, researchers in the field organized two MICCAI grand challenges to draw  attention on 6-month-old infant brain MRI segmentation: \href{http://iseg2017.web.unc.edu/}{iSeg-2017}\cite{wang2019benchmark} and iSeg-2019 \cite{sun2021multi-infant}.
In the first one, deep learning-based methods have shown their promising performance on 6-month-old infant subjects from a single site. 
The iSeg-2019 challenge \cite{sun2021multi-infant} instead addressed the segmentation of infant brain MRI from multiple centers. 

\subsection{Best methods from iSeg-17}

Characteristics and trade-offs of the top-performing submissions in the iSeg-2017 challenge are reported in  Table~\ref{tab:iseg17-methods} summarizing the key components of each method, including architecture type, training strategy, and inference time.

Among the top-performing methods, deep learning-based approaches dominated, with a clear focus on dense and fully convolutional architectures. Several works, such as \cite{bui20173d} and \cite{dolz2020deep}, extended densely connected networks to facilitate effective feature propagation and contextual learning, often using ensemble models and sub-volume sampling strategies to balance accuracy and computational cost. Other methods, like \cite{zeng2018multi} and \cite{moeskops2017isointense}, introduced multi-scale or hybrid 2D/3D frameworks. These leveraged skip connections, dilated convolutions, and patch-based training to preserve spatial resolution and integrate complementary contextual information.

Transfer learning also emerged as a promising approach: \cite{xu2017neonatal}, for instance, fine-tuned a VGG-based fully convolutional network pre-trained on ImageNet, achieving both high accuracy and exceptional inference speed. Similarly, \cite{fonov2018neuromtl} exploited a large external dataset (\href{https://nda.nih.gov/edit_collection.html?id=19}{IBIS}) for pre-training, improving robustness through fine-tuning and data augmentation. \cite{milletariVNetFullyConvolutional2016a} extended the V-Net architecture by introducing augmented paths and voxel masks to enhance resolution and precision.

Interestingly, \cite{sanroma2016building} proposed the only non-deep-learning approach among the top contenders, combining multi-atlas label fusion with an SVM classifier in a cascaded fashion. While this method achieved competitive segmentation quality, it came at the cost of significantly longer inference times.

Despite architectural diversity and training strategies, all methods experienced a notable drop in Dice coefficient when tested on data acquired from different sites or scanners, underscoring the persistent challenge of domain generalization in infant brain MRI segmentation.

\begin{table*}[ht]
\centering
\caption{Summary of top-performing methods from the iSeg-2017 challenge.}
\label{tab:iseg17-methods}
\begin{tabular}{p{4cm} p{3cm} p{5cm} p{2.3cm}}
\toprule
\textbf{Method} & \textbf{Architecture} & \textbf{Training Strategy} & \textbf{Inference Time} \\
\midrule
\cite{bui20173d} & Dual-path DenseNet & sub-volume samples, majority voting& $\sim$5 min \\
\cite{dolz2020deep} & SemiDenseNet & No augmentation, ensemble of 10 CNNs & $\sim$10 sec $\times$ 10 \\
\cite{zeng2018multi} & 2-stage 3D FCN & Multi-scale supervision, context information & $\sim$8 sec \\
\cite{moeskops2017isointense} & 2D + 3D dilated CNN & Patch-based training, no
data augmentation& $\sim$1 min \\
\cite{sanroma2016building} & Multi-atlas + SVM & Cascaded ensemble, non-rigid registration& $\sim$30 min \\
\cite{fonov2018neuromtl} & 3D U-Net & First training on \href{https://nda.nih.gov/edit_collection.html?id=19}{IBIS}, patch-based training & $\sim$8 sec \\
\cite{milletariVNetFullyConvolutional2016a} & Augmented V-Net & Patch-wise + ROI mask + augmentation& $\sim$6 sec \\
\cite{xu2017neonatal} & FCN based on
VGG16 & Transfer learning
(pre-trained on ImageNet)  & $\sim$1.8 sec \\
\bottomrule
\end{tabular}
\end{table*}

\subsection{Best methods from iSeg-19}

A summary of all top performing methods of the iSeg-2019 challenge is provided in Table~\ref{tab:iseg2019-summary}.
Participant teams had to reuse the same 10 training subjects from iSeg-2017, to enable a fair comparison across years and to evaluate improvements in segmentation algorithms over time. 
The main difference with the 2017 challenge is in the evaluation and benchmarking setup, as iSeg-2019 focuses more on cross-site generalization by using test data acquired from three independent sites: the University of North Carolina at Chapel Hill/University of Minnesota (Baby Connectome Project), Stanford University, and Emory University. This setup introduced domain shifts that made the task particularly challenging. 

The top-performing methods fell into three categories: attention-based models, U-Net variants, and domain adaptation techniques. Among the attention-based models, \cite{lei2019infant} proposed QL111111, a 3D U-Net with residual blocks and dilated convolutions. They applied contrast augmentation and self-attention, using leave-one-out cross-validation and majority voting. 
Within the challenge, the team led by Zhong et al. \cite{sun2021multi-infant} presented {Tao\_SMU}, a dual-stream attention-guided network with self- and pooling-attention modules, trained with contrast transformations and tested via sliding-window inference. Several methods were based on U-Net \cite{cicek3DUNetLearning2016b}:
Jun et al. introduced {FightAutism} \cite{sun2021multi-infant}, a 3D U-Net with histogram matching, instance normalization, and long skip connections. 
The method called {xflz} by Feng et al. \cite{sun2021multi-infant} extended U-Net with tissue-specific contrast augmentation and intensity-aware encoding blocks. 
Basnet et al. developed a ResDense U-Net \cite{sun2021multi-infant} using dense and residual connections, trained with Dice and cross-entropy loss. 
Domain adaptation played a key role due to site variability. 
Ma et al. proposed SmartDSP \cite{sun2021multi-infant}, built on nnU-Net \cite{isensee2021nnu} with adversarial learning and a bottleneck domain discriminator. 
Yu et al. introduced EMNet \cite{sun2021multi-infant}, a combined DenseNet-style network with entropy minimization and adversarial domain alignment. 
Finally, the team lead by Trung presented {Cross-linked FC-DenseNet} \cite{sun2021multi-infant}, featuring spatial and channel squeeze-and-excitation blocks \cite{roy2018concurrent} and cross-link connections, trained with random cropping. 
Although these approaches improved segmentation performance, domain shift remained a major barrier, and no method proved universally robust across acquisition sites.

\begin{table*}[ht]
\centering
\caption{Summary of top-performing methods from the iSeg-2019 challenge.}
\label{tab:iseg2019-methods}
\begin{tabular}{p{4cm} p{3cm} p{5cm} p{2.3cm}}
\toprule
\textbf{Method} & \textbf{Architecture} & \textbf{Training Strategy} & \textbf{Inference Time} \\
\midrule
\cite{lei2019infant} & 3D U-Net + attention & DCP blocks, self-attention, ensemble & $\sim$2 min \\
{Tao\_SMU} (Zhong et al.) & Attention-guided full-res net & Dual attention, multi-view augmentation & $\sim$12 min \\
{Fightautism} (Jun et al.) & 3D U-Net & Histogram matching, nnU-Net-like training & N/A \\
{xflz} (Feng et al.) & Optimized 3D U-Net & Tissue-intensity aug., noise/flip, patch-wise & $\sim$1 min \\
{SmartDSP} (Ma et al.) & nnU-Net + adv. DA & Feat-level discriminator, Dice+CE+focal loss & $\sim$10 sec \\
{EMNet} (Yu et al.) & DenseNet + EMNet & Entropy min., adversarial DA, IBN layers & $\sim$8 sec \\
Trung et al. & Cross-linked FC-DenseNet & scSE blocks, cross-links, dropout, patch-wise & $\sim$6 min \\
Basnet et al. & U-DenseResNet & Dense/residual blocks, dropout, Dice+CE loss & $\sim$1 min \\
\bottomrule
\label{tab:iseg2019-summary}
\end{tabular}
\end{table*}

\subsection{Other recent work}
Outside of the iSeg challenge publications, there have been other proposed deep-learning-based segmentation approaches in infants.
Among these, some of them exploit DenseNets and their potential of feature reuse for the task of brain segmentation.
For example, \cite{hashemi2019exclusive} presented a fully convolutional DenseNet, trained with a similarity loss function, for the segmentation of mutually exclusive brain tissues. 
Building on this idea, \cite{qamar2020variant} introduced an adapted U-Net that incorporates DenseNet for integrating low-level features along the encoding path and employs inceptionResNet for merging high-level features in the decoding path. 
In a related contribution, \cite{qamar2019multi} proposed a multi-path hyperdensely connected model, establishing dense connections between layers across various channels to facilitate information fusion at early, intermediate, and advanced stages of the network. 
Other works have explored alternative architectural strategies: for instance, \cite{ding2021multimodal} 
introduced a framework to combine fuzzy adjacency in a deep learning model trained with multimodal MRI scans.
The authors evaluated the architecture on the iSeg-17 publicly available dataset, validating the superiority of fuzzy guidance and deep supervision structures, with the only limitations of losing some global contextual representation due to operating on $32^3$ subvolumes rather than the entire volume. 
It is worth noting that patch-dependent techniques suffer from a limitation in preserving global context information, which is essential for ensuring the spatial consistency of tissue segmentation \cite{ReinaFrontiers2020}.

Recently, \cite{zeng20233d} constructed a 3D mixed-scale asymmetric segmentation network (3D-MASNet) framework by embedding a well-designed 3D mixed-scale asymmetric convolution block (MixACB) into existing segmentation CNNs.
Then, they evaluated the effectiveness of the MixACB on five canonical CNN networks using the iSeg-2019 training dataset. 
The MixACB significantly improved the segmentation accuracy of various CNNs, among which DU-Net \cite{wang2018volume} with MixACB achieved the best-enhanced average performance, ranking first in the iSeg-2019 Grand Challenge (Dice coefficients
of 0.931 in GM, 0.912 in WM, and 0.961 in CSF).
Another work that performed well on MR images from the iSeg-2019 challenge is \cite{sun2021multi}.
In the training stage, they first estimated coarse tissue probabilities and build a global anatomic guidance. 
They then trained another segmentation model based on the original images to estimate fine tissue probabilities. The global anatomic guidance and the fine tissue probabilities were integrated as inputs to train a final segmentation model. 
In the testing stage, they proposed an iterative self-supervised learning strategy to train a site-specific segmentation model based on a set of reliable training samples, which are automatically generated and iteratively updated, for a to-be-segmented site.

%%%%%%%%%%%%%%%%%%%%%%%%%%%%%%%%%%%%%%%%%%%%%%%%%%%%%%%%%%%%%%%
%%%%%%%%%%%%%%%%%%%%%%%%%%%%%%%%%%%%%%%%%%%%%%%%%%%%%%%%%%%%%%%

\section{Brain MRI data}
\label{sec:data}

In this study, we leverage multiple publicly available datasets of both adult and infant brain MRI. 
These datasets span a wide range of ages, imaging sites, and acquisition protocols, providing diverse inputs for the segmentation task.

Adult brain MRI data are sourced from a previous study \cite{svanera2024fighting}, which gathered, from 80 databases covering approximately 160 world sites, almost 27,000 adult brain T1-weighted volumes of both healthy and clinical subjects. 

Infant brain data mainly relies on the LifeSpan Baby Connectome Project (\href{https://www.developingconnectome.org/project/}{BCP}) and the developing Human Connectome Project (\href{https://www.developingconnectome.org/project/}{dHCP}) datasets.
The \href{https://www.developingconnectome.org/project/}{BCP} dataset originally consists of 1,869 MRI volumes collected from 349 participants aged from birth (0 months) to 6 years (80 months) at the University of North Carolina (UNC) and the University of Minnesota (UMN). 
The \href{https://www.developingconnectome.org/project/}{dHCP} dataset originally includes 707 neonatal MRI scans acquired at a single site (Evelina Newborn Imaging Centre, UK), covering the age range from 20 weeks gestation to term (44 weeks post-menstrual age), equivalent to 0--4 months from birth.
To further assess model robustness and generalizability, we incorporate additional external infant raw brain MRI datasets, including \href{https://brain-development.org/brain-atlases/neonatal-brain-atlases/neonatal-brain-atlas-gousias/}{ALBERTS}, \href{https://osf.io/4vthr/}{MCRIB2}.

Finally, in order to cope with skull-stripped infant data such those in the \href{https://iseg2019.web.unc.edu/}{iSeg-19} challenge (i.e., in which non-brain tissues, such as the skull, scalp, and dura, have been removed), we also include skull-
stripped volumes from the Baby Connectome Project (BCP) and iSeg-19 itself.
In Fig.~\ref{fig:infant-datasets}, we detail the infant data preparation process, the exclusion criteria, and present details of each dataset, including age range, utilized volumes, data splits, and ground-truth generation.

\subsection{Adult Brain Training Data}

From \cite{svanera2024fighting}, we exploit for training a randomised selection of 1,049 volumes from 70 different sites (15 volumes for each dataset, except one contributing with 14 volumes).
The validation set, used for hyperparameter selection, includes 117 volumes from 72 datasets (91 internal and 26 external).

The labelling strategy for adult brains follows the 7 classes adopted by MRBrainS challenge \cite{mendrikMRBrainSChallengeOnline2015a}: \textit{grey matter, white matter, cerebrospinal fluid, ventricles, cerebellum, brainstem, and basal ganglia}.
Such labeling maximises the possibility of comparison with other state-of-the-art methods, and covers most of clinical and research studies and applications.
As no manual segmentations (gold standard) are available for any volume, we exploit FreeSurfer \cite{fischlFreeSurfer2012a} to produce a silver standard ground-truth (i.e., labels produced by a widely accepted method, with errors). 
Further information on adult training data can be retrieved from \cite{svanera2024fighting}.

\subsection{Raw Infant Brain Data}

\begin{figure*}[h]
     \centering
     %\begin{subfigure}[b]{\columnwidth}
         \centering
         \includegraphics[width=\textwidth]{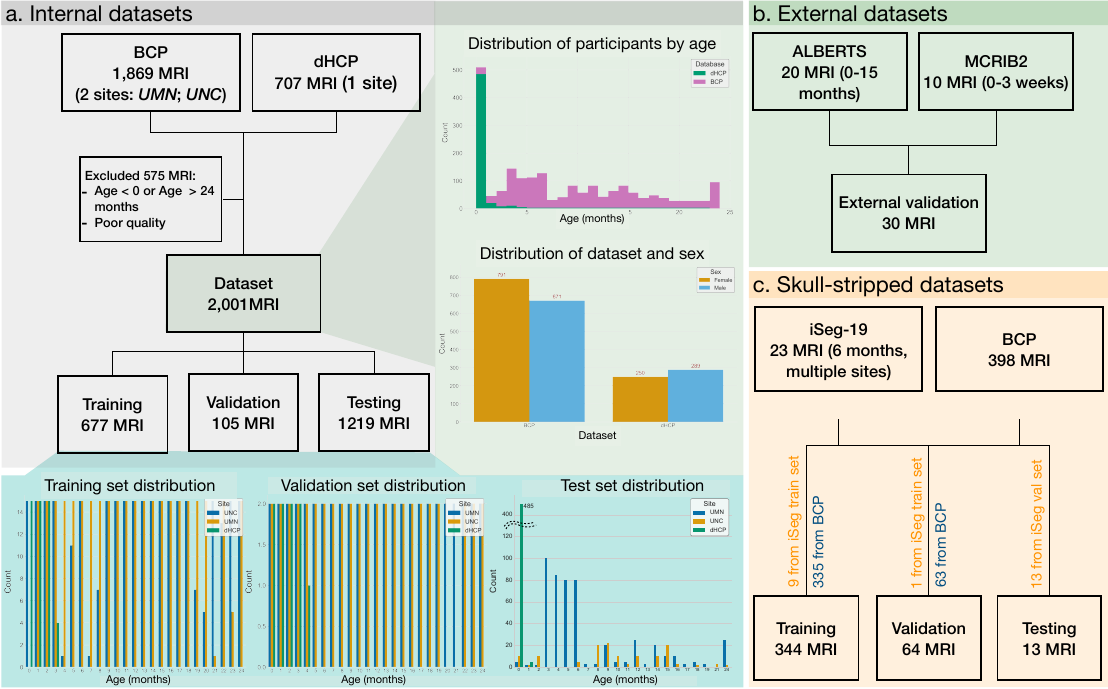}
        \caption{Infant brain MRI data: (a) Internal Datasets Flow Diagram for Inclusion and Exclusion of dataset and their distributions; (b-c) Flow Diagrams for External, and Skull-stripped datasets.}
        \label{fig:infant-datasets}
\end{figure*}

Although the World Health Organization (WHO) defines the infant period as spanning from 28 days to 1 year of age, in this work we adopt a broader operational definition, considering brain MRI scans from subjects aged 0 to 2 years.
Scans from premature infants and low-quality volumes are excluded.
To maintain sequence data consistency between adult and infant brain images, T1w whole-head infant  (i.e., including the skull, scalp, and other non-brain tissues) MRIs scans are retained.
Ultimately, 2,001 raw infant brain volumes from \href{https://www.developingconnectome.org/project/}{BCP} and \href{https://www.developingconnectome.org/project/}{dHCP} 
 datasets pass the quality control process.
Fig.~\ref{fig:infant-datasets} illustrates the main properties of these volumes.
 
The pre-processing pipeline of infant brain involves resampling each image to a standard isotropic voxel size of $1mm^3$ and reorienting it to a common anatomical space (RAS+orientation), where the origin is aligned with the anterior commissure, and axes correspond to right-left, anterior-posterior, and superior-inferior directions. 
%Additionally, image data are converted to a standard floating-point format, and cubic interpolation is applied during resampling. 

For training, validation, and testing, the dataset is split as shown in Fig.~\ref{fig:infant-datasets}: the training set includes up to 15 volumes per age group per site, resulting in 677 volumes. 
The validation set consists of a maximum of two volumes per age group per site, totaling 105 volumes. 
The remaining 1,219 volumes are used for testing.

The labelling strategy is the same adopted for adult brains  (grey matter, white matter, cerebrospinal fluid, ventricles, cerebellum, brainstem, and basal ganglia).
The silver ground-truth (GT) masks are  generated using Infant FreeSurfer \cite{zollei2020infant}, which is the dedicated infant brain processing pipeline developed as the counterpart to the standard FreeSurfer, designed for early postnatal brain anatomy and distinguished primarily by its use of an age-appropriate atlas.

\subsection{External Datasets of Raw Infant Brains}
The \href{https://brain-development.org/brain-atlases/neonatal-brain-atlases/neonatal-brain-atlas-gousias/}{ALBERTS} dataset includes 20 MRI volumes from infants aged 0 to 15 months, all of which are used as external test data without exclusions. Similarly, the \href{https://osf.io/4vthr/}{MCRIB2} dataset comprises 10 neonatal MRI volumes from infants aged 0 to 3 weeks, and all are included in the external evaluation set.
All GT labels are generated using Infant FreeSurfer.

\subsection{Skull-stripped Infant Brains}

To develop a dedicated pipeline tailored for skull-stripped infant brain MRIs, we utilize both the \href{https://iseg2019.web.unc.edu/}{iSeg-2019} training set (10 volumes with gold standard manual GT on four classes - CSF, GM, WM, background -, which we split into 9 training and 1 validation volume for our purposes) and the iSeg-2019 validation set (13 volumes without GT, which we use as test volumes for a qualitative visual comparison against state-of-the-art tools).
Additionally, we incorporate 335 training and 63 validation skull-stripped volumes from the Baby Connectome Project (BCP), with GT labels (on four classes) generated using Infant FreeSurfer.

\section{Methods}
\label{sec:methods}

\subsection{Network Architecture}
\label{ssec:methods}

%%%

\begin{figure*}[h]
     \centering
     %\begin{subfigure}[b]{\columnwidth}
         \centering
    \includegraphics[width=\textwidth]{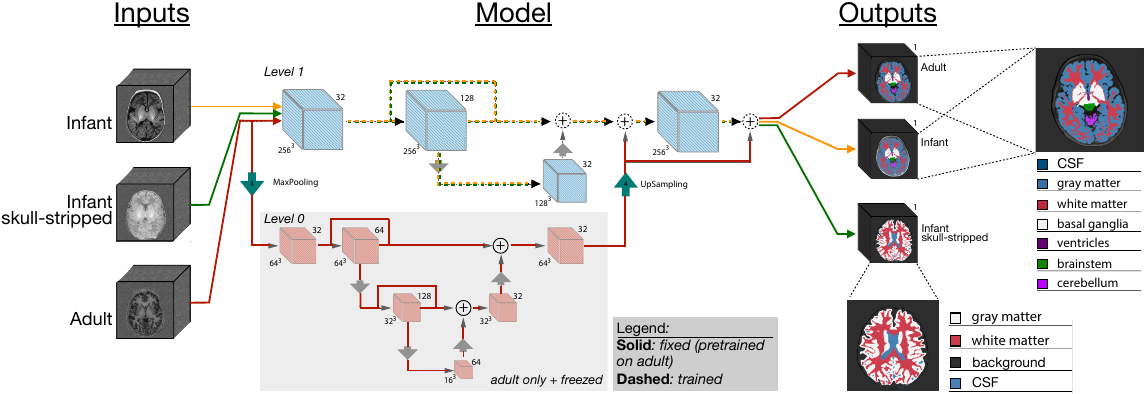}
        \caption{Proposed hierarchical 3D convolutional network for infant brain segmentation. The model follows a two-stage bottom-up training strategy: the lower-level network (\textit{grey}) learns a generalised anatomical prior from adult MRIs, while the upper-level network refines segmentation on infant brain images, adapting to age-specific anatomical variations.}
        \label{fig:model}
\end{figure*}

The proposed LODi model is a Level-of-Detail (LOD) network with a two-level structure designed for 3D volumetric brain data \cite{svanera2024fighting}.
As shown in Fig.~\ref{fig:model}, the network takes inputs of shape $(256 \times 256 \times 256)$ and produces a seven-class segmentation map of shape $(256 \times 256 \times 256 \times 7)$ for the raw infant brain model, and a four-class segmentation map of shape $(256 \times 256 \times 256 \times 4)$ for the skull-stripped variant.

At Level 0, the lower gray block in Fig.~\ref{fig:model}, the input is first downsampled using a MaxPooling operation to enable processing at lower resolution. 
The downsampled volume is then passed through a 3D convolutional layer (Conv3D) with 32 filters, followed by a series of convolutional blocks with 64 filters. The encoding path at this level concludes with a further downsampling step (with reduction factor $d = 4$ along each spatial dimension).
In the decoding path, the feature maps are progressively upsampled using UpSampling3D layers, with corresponding skip connections implemented via element-wise addition (Add layers) to recover spatial detail, support high-level feature fusion, and perform anatomical localization.

At Level 1, the upper blocks in Fig.~\ref{fig:model}, the higher-level features from Level 0 are further refined through additional convolutional blocks with 128 filters, enabling to refine segmentation masks at finer scales. The decoder at this level mirrors the encoder?s structure, again using UpSampling3D layers and Add-based skip connections.
ReLU is used as the non-linear activation function throughout both levels of the encoder and decoder. 
Group Normalization (GN) is applied for improved stability, with \cite{he2015delving} initialization ensuring efficient weight initialization. 
A dropout rate of 0.05 is incorporated to mitigate overfitting. 
Additionally, skip pathways and cross-level summation nodes improve segmentation accuracy while maintaining parameter efficiency, and represent a more parameter-efficient solution compared to concatenation-based strategies \cite{milletariVNetFullyConvolutional2016a}.
Ultimately, the network is highly lightweight (with only 337,719 trainable parameters for the adult brain model), orders of magnitude fewer than competing architecture like SynthSeg \cite{billot2021synthseg}, which exceed 15 millions.

\subsection{Raw Infant Brain Model: Training}

The training process follows a two-stage bottom-up strategy. 
Initially, the network hierarchy is trained on adult brain MRIs only, as in \cite{svanera2024fighting}.
The lower-level network (in gray color in Fig.~\ref{fig:model}) establishes an \textit{anatomical prior} based on adult training data. 
Since reducing the input resolution eliminates fine-grained structural details and attenuates subject-specific variations, this stage learns a generalized representation of adult brain anatomy and its spatial organization \cite{svanera2024fighting}. 
This abstraction ensures robustness against variations introduced by different acquisition sites and anatomical differences.
The loss function used at this stage is a pure per-channel Dice loss (averaged across labels).
More details can be found in \cite{svanera2024fighting}.

Once the lower level reaches convergence, its parameters are frozen, and cross-level connections propagate the spatial context learned from adult brains to the upper level (light blue blocks in Fig.~\ref{fig:model}), which is trained from scratch with infant data using per-channel Dice as loss function. 
Building upon the anatomical foundation of adult brains, the final adaptation stage trained with infant brain data introduces the flexibility needed to adapt to age-specific anatomical variations, refining the model to better accommodate the structural differences characteristic of early brain development.

Training employs the Adam algorithm \cite{kingma2014adam}, with the training duration set to $100$ epochs for the upper level with frozen lower level. 
The initial learning rate is $5e-4$, which is decreased by a factor of $1/4$ when a plateau is reached. 
The training process takes $\sim24h$ on a workstation equipped with Nvidia$\textcopyright$ A100-SXM4-80GB, with experiment tracking managed through \href{https://www.wandb.com/}{\textit{Weights \& Biases}}.

\subsection{Skull-Stripped Infant Brain Model: Training}

We also develop a skull-stripped infant brain segmentation model designed to operate on four-class segmentation maps (gray matter, white matter, cerebrospinal fluid, and background), in line with widely adopted benchmarks for this type of data (e.g., iSeg challenges). 
To align with a four-class segmentation task, we first replace the original seven-class output layer with a four-class output layer. The training procedure for the skull-stripped MRI model follows a three-stage bottom-up strategy, incorporating key adaptations to optimize performance on skull-stripped infant MRIs.

In the initial stage, we initialize the lower-level layers using weights pretrained on adult brain data, as described in \cite{svanera2024fighting}. 
In the second stage, we train the upper-level layers from scratch using 335 training and 63 validation skull-stripped MRI volumes from the BCP dataset, covering infants aged 4 to 8 months, labelled with Infant FreeSurfer. The optimization protocol remains consistent with that used for the full-head model, and training proceeds for 100 epochs to refine segmentation performance. 
Finally, in the third stage, we fine-tune the resulting model on 10 labeled skull-stripped MRIs from the iSeg-2019 training set (manual gold standard GT).

\subsection{Data augmentation}

\begin{table}[ht]
    \small
    \centering
%    \resizebox{\columnwidth}{!}{%
%    \begin{tabular}{l l l l}
    \begin{tabular}{p{1.5cm} p{2cm} p{0.5cm} p{2cm}}
        \toprule
        Group & Augmentation & Prob & Parameters \\
        \midrule
        \multirow{3}{*}{Geometrical} 
            & Translation    & 1/3 & 3 axes; shift $\pm$20 voxels; zero padded \\
            & Rotation       & 1/3 & 3 axes; $\pm$10 degrees; zero padded \\
            & Grid distortion& 1/3 & Steps: 4; Distortion: 0.1 \\
        \midrule
        \multirow{6}{*}{Noise}
            & Blur           & 1/9 & Limit: 3 \\
            & Salt and pepper& 1/9 & Amount: 0.01; Salt: 0.2 \\
            & Gaussian       & 1/9 & Amount: 0.2 \\
            & Downscale      & 1/9 & Scale: 0.25--0.75 \\
            & Gamma          & 1/9 & Clip: 0.025 \\
            & Contrast       & 1/9 & Alpha: 0.5--3.0 \\
        \midrule
        \multirow{4}{*}{Artefacts}
            & Ghosting       & 1/9 & Max rep.: 4 \\
            & Slice Spacing  & 1/9 & Axial plane only; Spacing: 2--5mm \\
            & Inhomogeneity  & 1        & See \cite{svanera2021cerebrum} for details \\
            & Field Bias     & 1/9 & Num cycles: 5; Scale factor: 2 \\
        \bottomrule
    \end{tabular}
%    }
    \caption{Augmentation methods with their probabilities and parameters. We use blur, downscale, and grid distortion from \cite{info11020125}, and ghosting from \cite{perez-garcia_torchio_2021}. The intensity-based augmentations are subdivided into two groups: noise and artefacts.}
    \label{tab:augmentations}
\end{table}

During model training of both pipelines (raw infant and skull-stripped brain data), we employ a data augmentation pipeline on the 90\% of volumes. 
Table \ref{tab:augmentations} summarizes the applied augmentations and their parameters. 
These augmentations fall into two main categories: geometric and intensity-based. Geometric augmentations, such as translation and rotation, modify the positioning and alignment of the brain volume (and the segmentation mask) within the voxel grid. Intensity augmentations, including blurring, noise addition, and contrast adjustments, alter the perceived voxel intensities. The probability values and augmentation parameters shown in Table~\ref{tab:augmentations} are empirically selected to optimize performance and maintain data integrity.

%%%%%
\section{Results and Discussion}
\label{sec:results}

The experimental assessment of our model for infant brain segmentation is based on  quantitative and qualitative comparisons of our method against the state-of-the-art.
Quantitative results are computed using Dice coefficient as performance metric, and using Infant FreeSurfer segmentation as silver ground-truth, unless differently stated.

%%%%%%%%%%%%%%%%%%%%%%%%%%%%%%%%%%%%%%%%%%%%%%%%%%%%%%%%%%%%%%%
\subsection{Quantitative method comparisons on raw infant data}
\label{ssec:soa-comparison}

We present a comparative evaluation of our proposed method LODi against state-of-the-art brain segmentation techniques, focusing on segmentation accuracy across different brain structures. 
The benchmark methods were selected based on their performance on infant brain MRI data, ensuring a fair and meaningful assessment. 
While several popular brain segmentation methods exist, many of them perform poorly when applied to infant data.
For this reason, after the initial performance assessment, we decided to exclude CAT12 from SPM \cite{gaser2022cat},
FSL-BET \cite{smith2002fast}, and FreeSurfer \cite{fischlFreeSurfer2012a}. 
Among the available alternatives, only SynthSeg \cite{billot2021synthseg} and LOD-Brain \cite{svanera2024fighting} demonstrated satisfactory performance and were thus included. 
\textit{SynthSeg}, a leading synthetic data-driven approach, leverages large-scale synthetic training to generalize across heterogeneous MRI scans. 
\textit{LOD-Brain} represents one of the most recent and robust 3D segmentation pipelines developed for adult brain MRI.
In addition to these methods, we also evaluate a version of LODi trained completely \textit{from scratch}, that is when both the lower and upper levels were trained without using pretrained weights obtained on adult data. 

\begin{figure}[h!]
    \centering
     \begin{subfigure}[b]{\columnwidth}
         \centering
         \includegraphics[width=\textwidth]{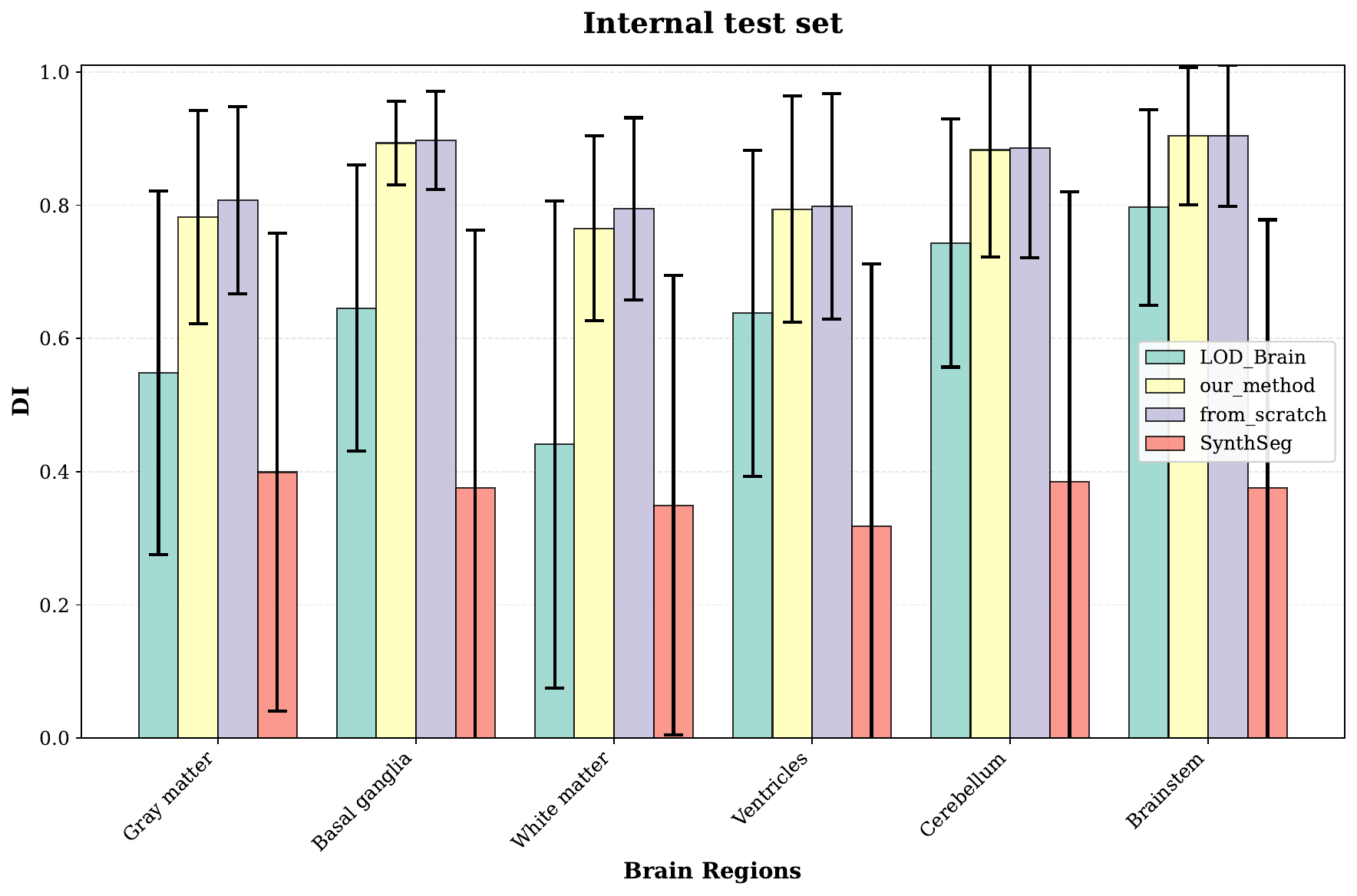}  % fixed "basal ganglia"
         \caption{Internal datasets only with results grouped by brain structure.}
         \label{fig:model_comparisons_INT}
     \end{subfigure}
     \begin{subfigure}[b]{\columnwidth}
        \vspace{1em}
         \centering
         \includegraphics[width=\textwidth]{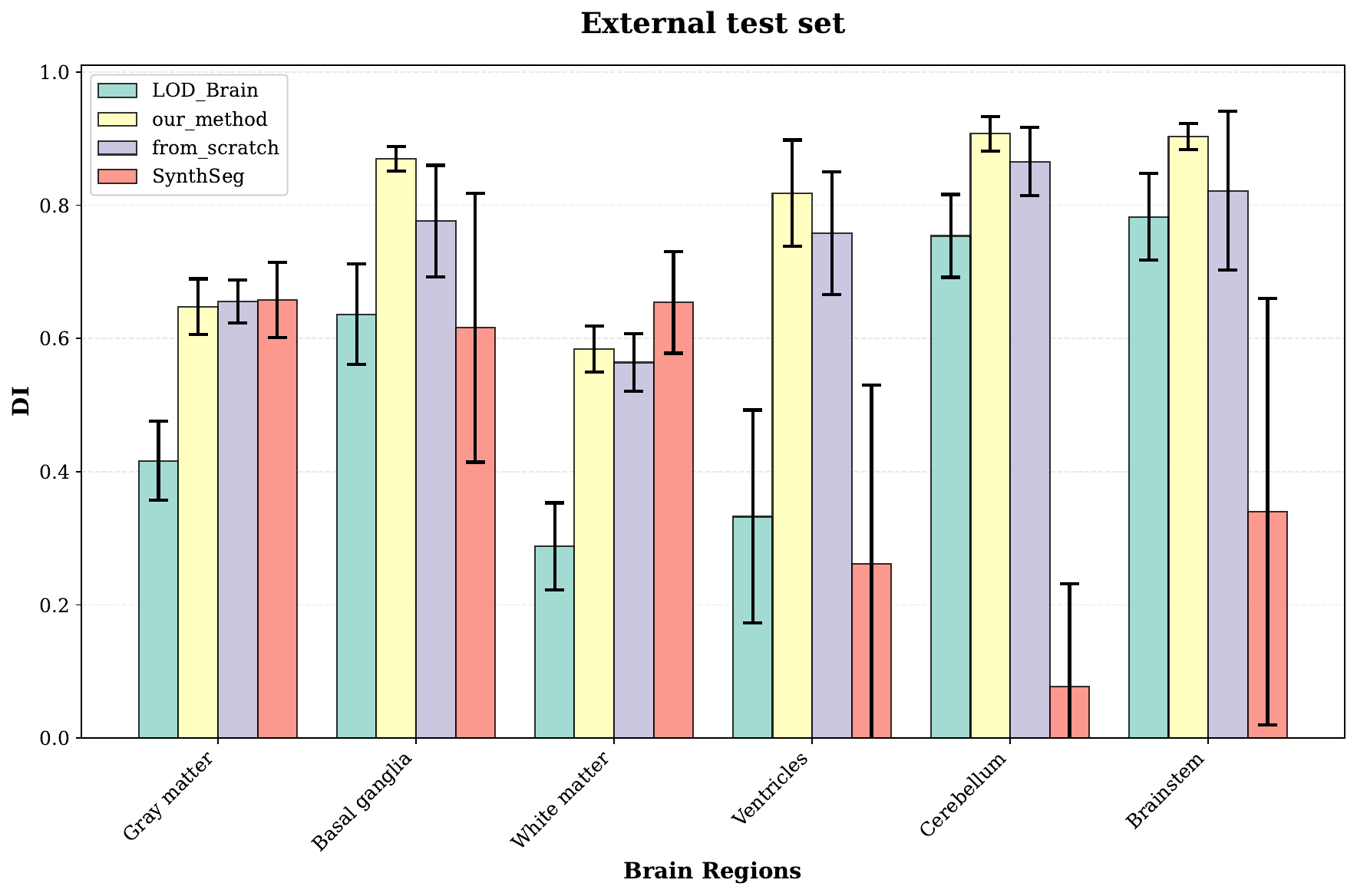}
         \caption{External datasets only with results grouped by brain structure}
         \label{fig:model_comparisons_EXT}
     \end{subfigure}
         \caption{
    Performance comparison: SynthSeg \cite{billot2021synthseg}, LOD-Brain \cite{svanera2024fighting}, and our method LODi (trained also from scratch).
    (a) Results on the internal dataset are computed on the test set of $1,219$ volumes from BCP and dHCP, using Infant FreeSurfer as GT reference and grouped for brain structure.
    (b) Results on 30 volumes belonging to external sites (MCRIB2 and ALBERTS).
    }
    \label{fig:model_comparisons}
\end{figure}

All methods are evaluated against a silver standard ground truth provided by Infant FreeSurfer, and their prediction performance is assessed on selected test sets. 
Fig.~\ref{fig:model_comparisons_INT} illustrates segmentation results on the \textit{internal} testing sets, composed of 1,219 volumes from the BCP and dHCP datasets. 
Metrics are reported per brain structure, showing mean and standard deviation for each class, enabling a fine-grained analysis of model performance across anatomical regions. 
To further evaluate generalizability, Fig.~\ref{fig:model_comparisons_EXT} reports results on \textit{external} test sets-specifically MCRIB2 and ALBERTS-which include 30 volumes in total. 
Results confirm that the proposed models outperform competing methods on nearly all brain structures and achieves state-of-the-art accuracy on both internal and external datasets.
In particular, LODi demonstrates the benefits of transfer learning from adult brain segmentation, yielding a version of the model that is more robust than the same model trained from scratch (with infant brains only) on external datasets, thereby clearly enhancing generalization performance.

\subsubsection{Robustness to motion artifacts}

\begin{figure}[h]
     \centering
   \includegraphics[width=\columnwidth, trim=50pt 120pt 40pt 70pt, clip]{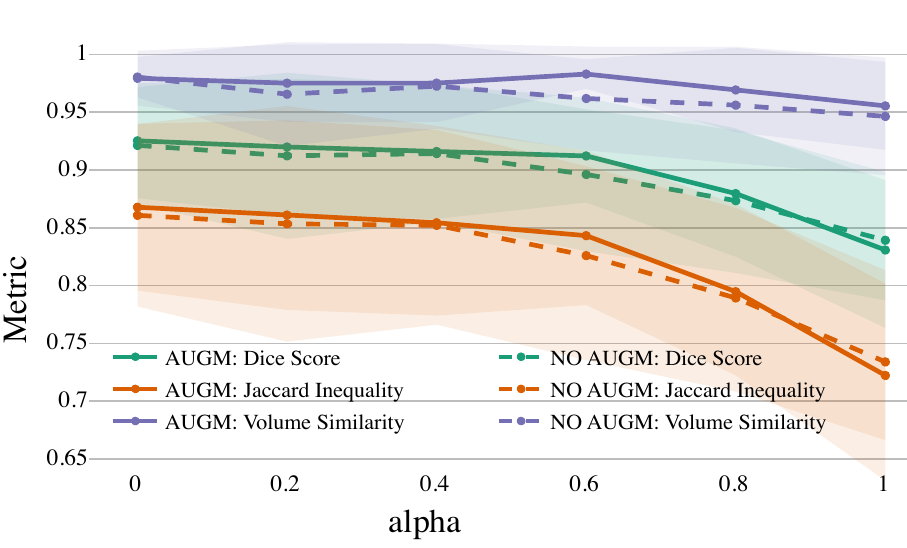}
        \caption{Robustness at different levels of motion artifacts.}
        \label{fig:display_motion_artefacts}
\end{figure}

During the image acquisition process, for this age range the quality of MRI images can be compromised by motion artifacts, which may adversely affect clinical diagnoses and automated image analysis. 
Therefore, ensuring the robustness of a segmentation method to motion artifacts is crucial.
To assess the effectiveness of our approach in handling motion artifacts, we employ a method proposed by \cite{shaw2020k}, which allows for the generation of realistic motion artifacts on existing MRI data. 
Specifically, we apply this technique to 120 randomly selected testing volumes, with increasing values of $\alpha$ i.e., a parameter which controls the severity of simulated motion artefacts in MRI data. 
This parameter scales the magnitude of the randomly generated rigid 3D affine transformations applied to artefact-free MRI volumes. 
By adjusting $\alpha$, we simulate varying degrees of patient movement, thereby creating a range of motion artefacts from mild to severe.
In Fig.\ref{fig:display_motion_artefacts}, LODi demonstrate significant robustness to increasing values of motion artifacts.

\subsection{Qualitative comparisons on raw infant data}
\label{ssec:qualitative}

\begin{figure}[!h]
    \centering
    \includegraphics[width=1.0\columnwidth]{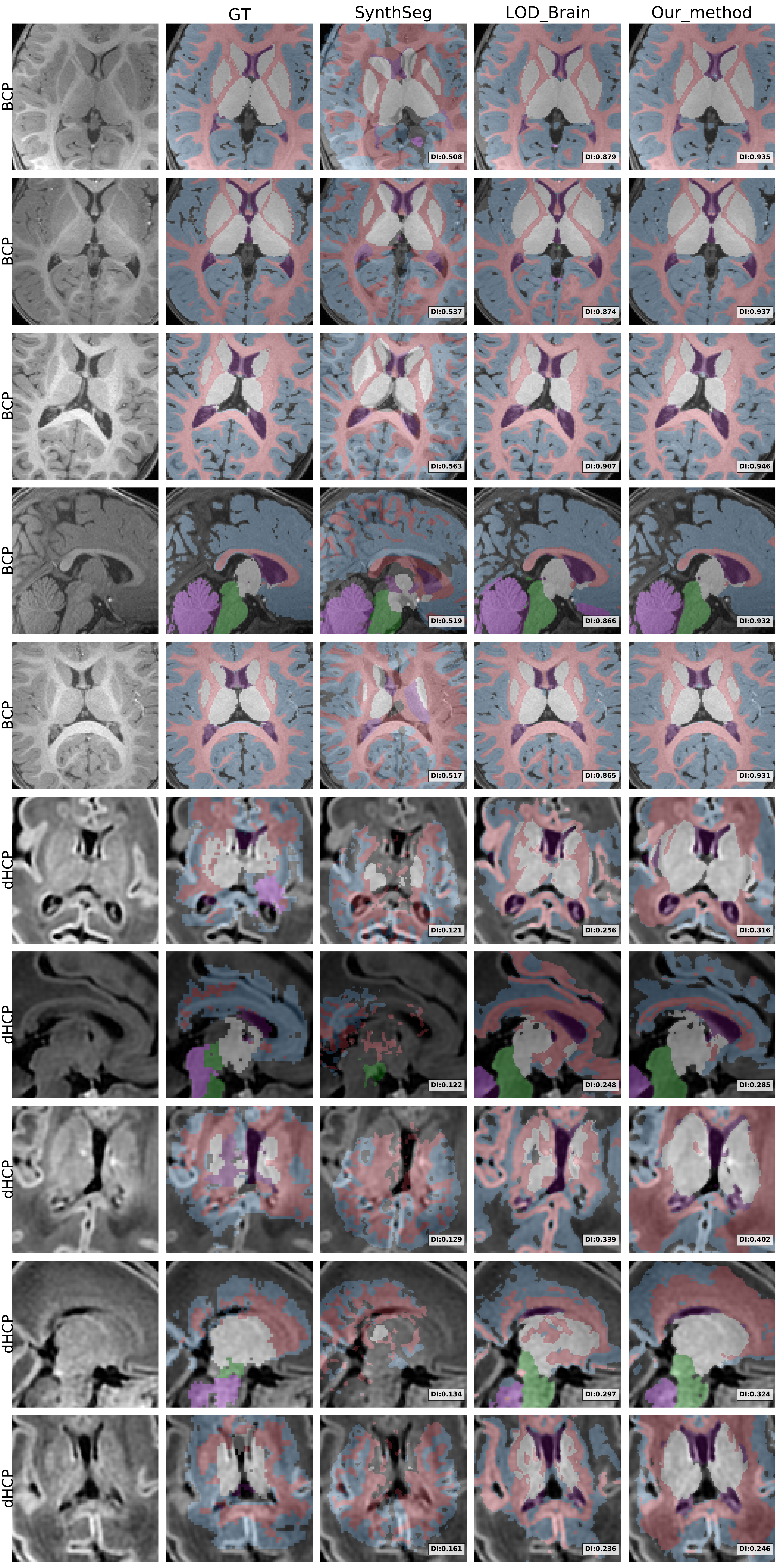}
    \caption{Comparison on the first half of 20 volumes (vol. 1-10) on the internal test set with highest disagreement on segmentation masks among difference methods. Zoom in for better view.}
    %\hl{FAREI 2 PAGINE/FIGURE}
    \label{fig:visual_results_method_comparison_a}
\end{figure}

\begin{figure}[!h]
    \centering
    \includegraphics[width=1.0\columnwidth]{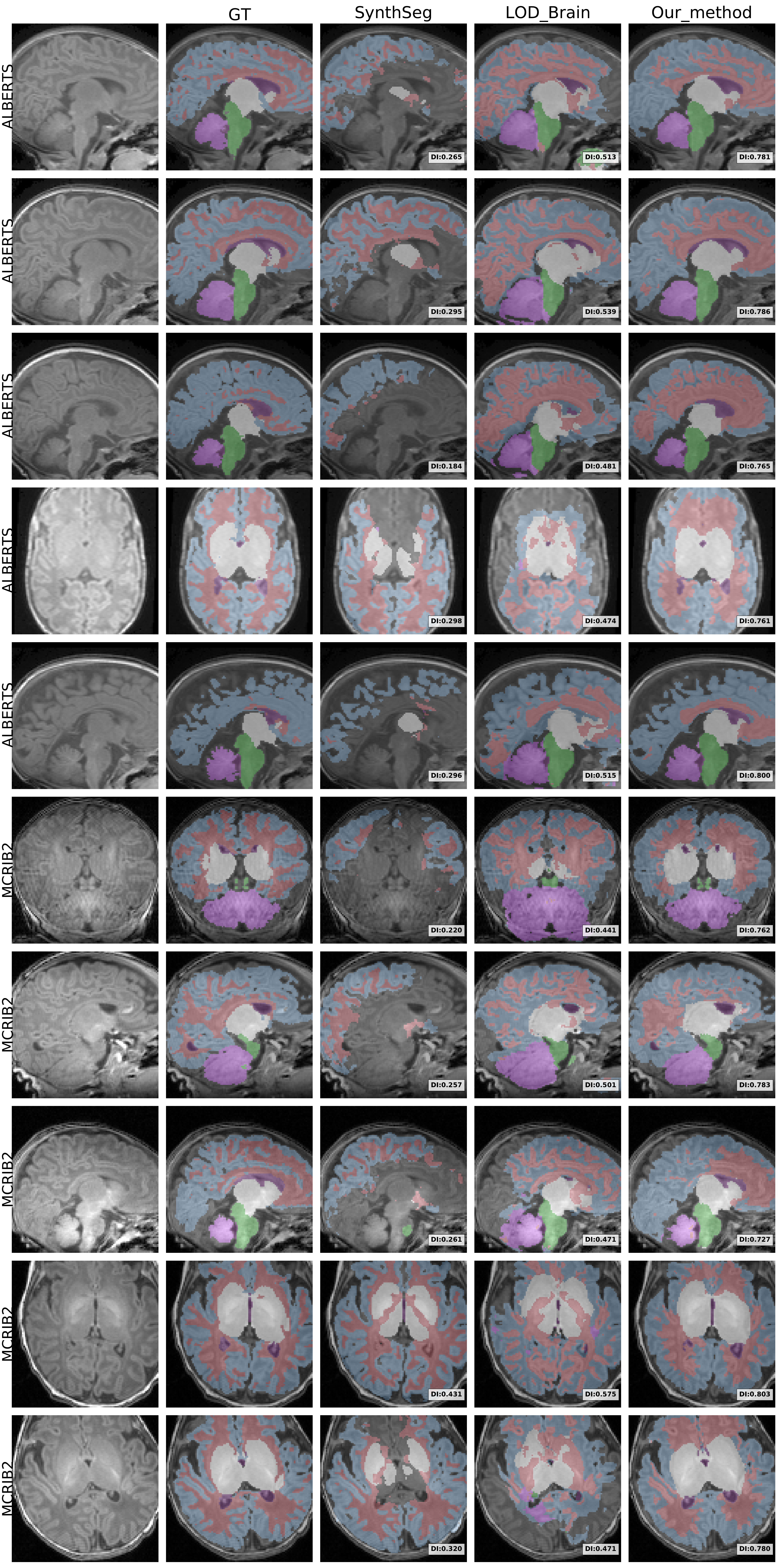}
    \caption{Comparison on the second half of 20 volumes (vol. 11-20) on the external test set with highest disagreement on segmentation masks among difference methods. Zoom in for better view.}
    %\hl{FAREI 2 PAGINE/FIGURE}}
    \label{fig:visual_results_method_comparison_b}
\end{figure}

To provide a clear visual assessment of segmentation performance, we present a diverse set of qualitative results obtained using different methods. Figs~\ref{fig:visual_results_method_comparison_a} and \ref{fig:visual_results_method_comparison_b} illustrate a side-by-side comparison between the competing approaches and the silver-standard ground truth generated by Infant FreeSurfer.
Specifically, we highlight the 30 most \textit{discordant} MRI volumes, identified as those exhibiting the highest variance in DICE scores across the evaluated methods when compared to Infant FreeSurfer segmentations. By focusing on these challenging cases, we enhance the visibility of key differences in anatomical structure delineation, allowing for a more insightful comparison of method effectiveness.

This visualization strategy ensures that performance disparities between compared approaches are clearly distinguishable, providing an intuitive and informative perspective on the strengths and limitations of each method, including our proposed solution.


\subsubsection{Surface analysis on raw infant data}
\label{ssec:surface-analysis}

\begin{figure*}[]
     \centering
     \begin{subfigure}[b]{1.0\columnwidth}
         \centering
         \includegraphics[width=1.\columnwidth]{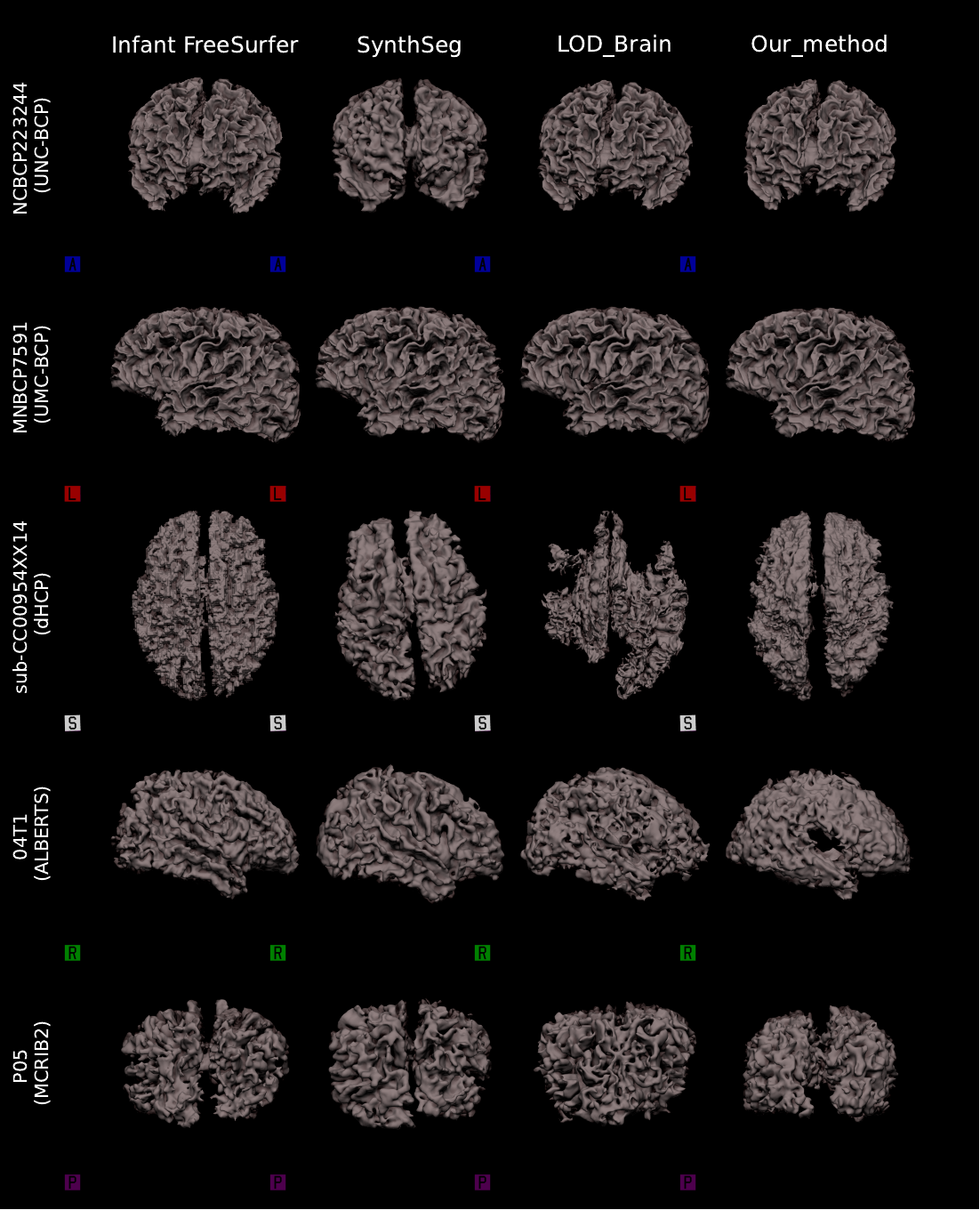}
         \caption{Inner Gray Matter surfaces}
         \label{fig:mesh_comparisons_WM}
     \end{subfigure}
\begin{subfigure}[b]{1.0\columnwidth}
        \vspace{3em}
         \centering
         \includegraphics[width=1.0\columnwidth]{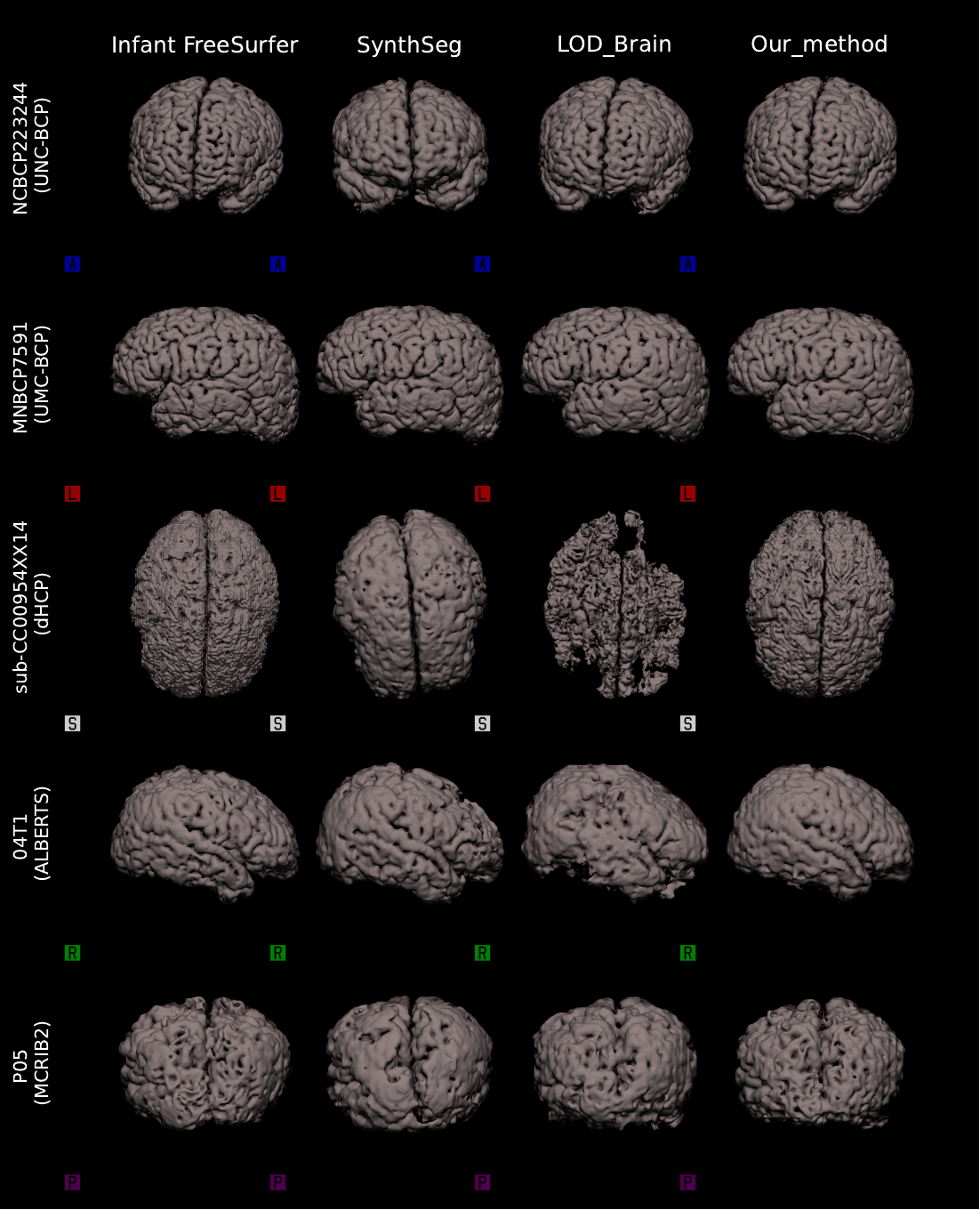}
        \caption{Outer Gray Matter surfaces}
         \label{fig:mesh_comparisons_GM}
        \end{subfigure}
    \caption{(a) Inner gray matter and (b) outer gray matter surfaces shown for the five test volumes exhibiting the highest DICE coefficient variance - one subject from each testing set}
     \label{fig:mesh_comparisons}
\end{figure*}

We perform a surface-based analysis using all methods, that are Infant FreeSurfer, SynthSeg, LOD-Brain, and LODi, to evaluate segmentation performance across different anatomical surfaces. 
Specifically, we analyze the surfaces of the inner gray matter (Fig.\ref{fig:mesh_comparisons_WM}), and of the outer gray matter (Fig.\ref{fig:mesh_comparisons_GM}).

This analysis is conducted on five MRI volumes, each selected from a different dataset (BCP-UMN-site, BCP-UNC-site, dHCP, ALBERTS, and MCRIB2), based on the scan exhibiting the highest variance in Dice scores across methods. Similar to the qualitative comparison, this selection strategy emphasizes cases where segmentation results show the greatest discrepancies.

By visualizing these 3D surface representations, generated using \href{https://github.com/neurolabusc/nii2mesh}{nii2mesh}, we provide a more detailed assessment of LODi?s ability to accurately capture anatomical brain structures. 
Notably, this approach also highlights the smoothness of the resulting segmentations, and facilitates a direct comparison with benchmark methods, offering deeper insights into segmentation performance across different surfaces.

\subsection{Evaluation on skull-stripped infant data}
\label{ssec:iseg-challenge}

\begin{figure*}[!ht]
  \centering
  \begin{minipage}[t]{0.45\textwidth}
    \centering
    \includegraphics[width=1.0\columnwidth,valign=t]{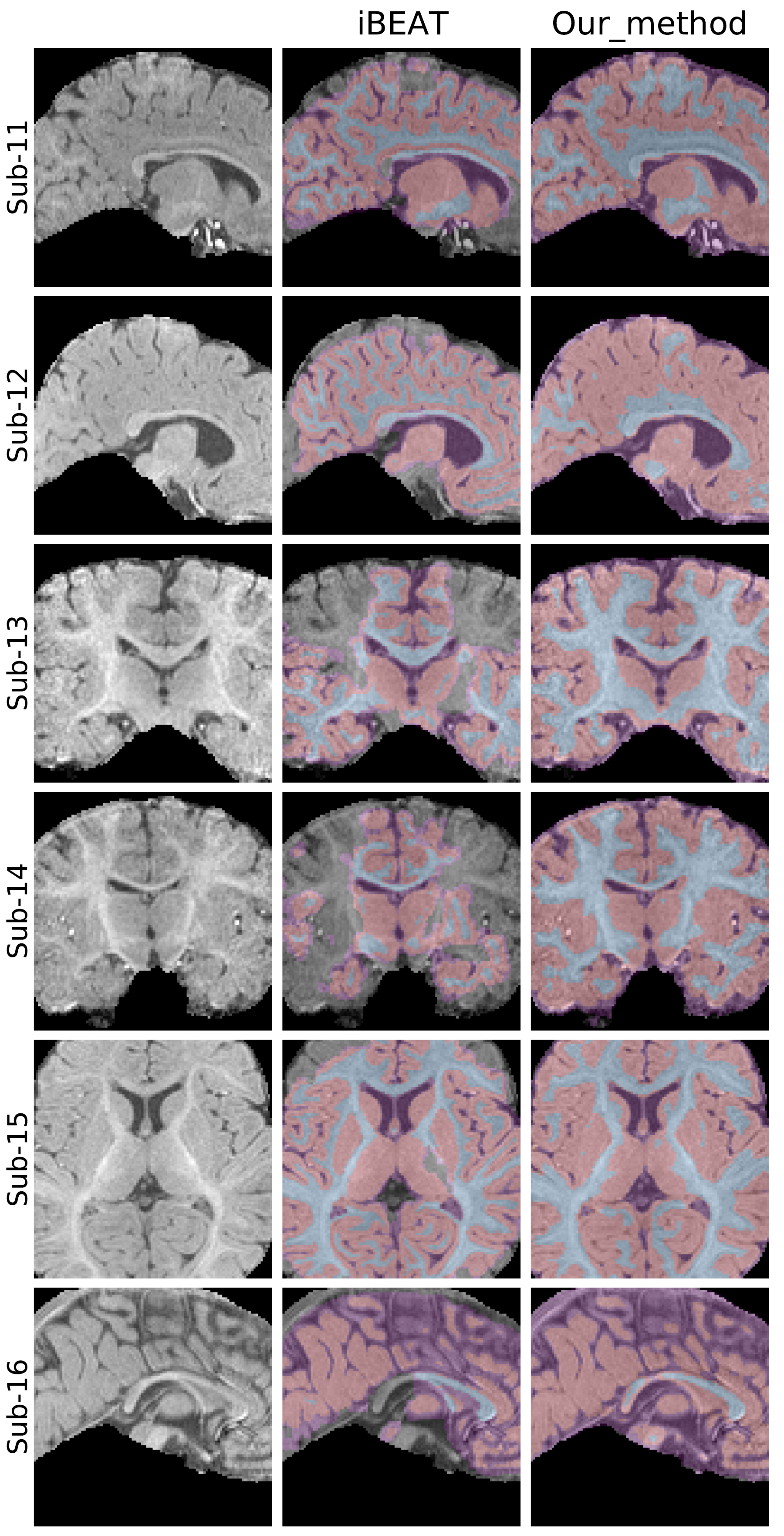}
    \vspace{0.2cm} % Adjust this space to place the (a) label correctly
    \centerline{(a)}
  \end{minipage}%
  \hspace{0.05\textwidth}
  \begin{minipage}[t]{0.45\textwidth}
    \centering
    \includegraphics[width=1.0\columnwidth,valign=t]{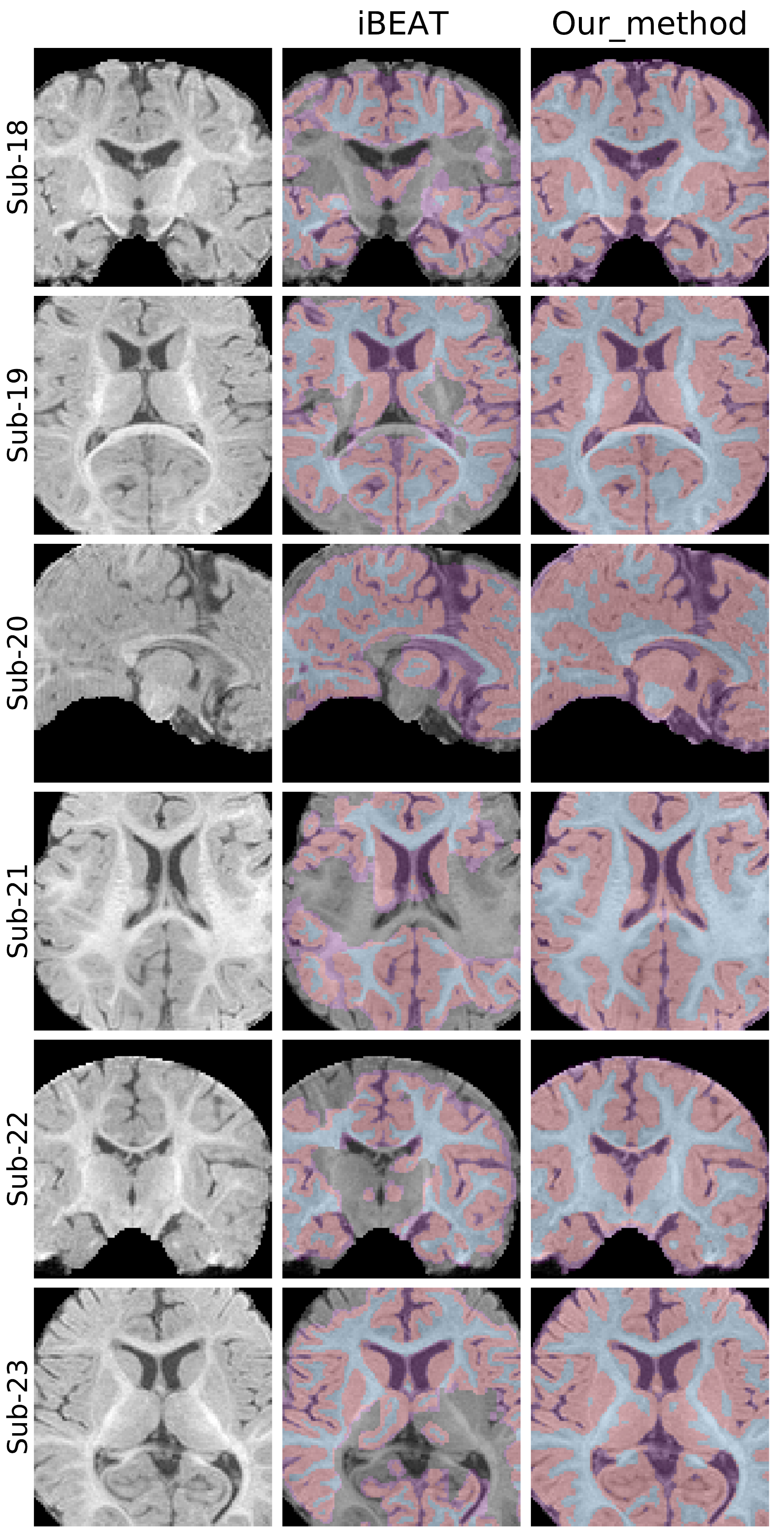}
    \vspace{0.2cm} % Adjust this space to place the (b) label correctly
    \centerline{(b)}
    \caption{Visual comparison between our method and iBEAT on the iSeg-19 test set: (a) Subjects 11-17, (b) Subjects 18-23.}
    \label{fig:iseg-comparison}
  \end{minipage}
\end{figure*}

Due to the absence of ground-truth labels for the iSeg-19 validation set, we assess the performance of LODi through a qualitative comparison against the \href{http://www.ibeat.cloud}{iBEAT V2.0 Cloud} \cite{wang2023ibeat} model.
iBEAT is a widely used infant brain segmentation framework which performs brain tissue segmentation into four primary classes: Gray Matter (GM), White Matter (WM), Cerebrospinal Fluid (CSF), and background. 
Fig.~\ref{fig:iseg-comparison} presents a side-by-side visual comparison of segmentation outputs from our model and iBEAT, focusing on the four-label segmentation task, and highlighting differences in the delineation of key brain structures. 
This qualitative analysis provides insights into the model's ability to generalize to unseen iSeg-19 data and maintain anatomical consistency across different segmentation approaches.
By visually inspecting the segmentations, we observe that our model produces more accurate results than iBEAT, demonstrating effective adaptation to skull-stripped infant MRI scans and the model's potential for reliable infant brain segmentation.

%%%%%%%%%%%%%%%%%%%%%%%%%%%%%%%%%%%%%%%%%%%%%%%%%%%%%%%%%%%%%%%

\section{Conclusion}
\label{sec:conclusion}

In this work, we introduce LODi, a novel infant brain segmentation framework that leverages adult brain priors to enhance segmentation accuracy in early neuro-developmental stages. 
By employing a two-stage bottom-up training strategy, our model effectively transfers anatomical knowledge from adult MRI scans to infant brain segmentation, mitigating the challenges posed by domain shifts, limited annotated infant data, and scanner variability.
Furthermore, we demonstrate the effectiveness of our approach through extensive experiments on internal and external datasets.
Additionally, by fine-tuning on skull-stripped MRI scans, we ensured improved generalization to datasets that follow different preprocessing protocols, including the iSeg-19 challenge volumes, where our model shows strong qualitative performance against the iBEAT framework.
Our findings highlight the potential of integrating developmental priors into deep learning-based neuroimaging models, paving the way for age-adaptive segmentation strategies that can generalize across different infant populations. 

%%%%%%%%%%%%%%%%%%%%%%%%%%%%%%%%%%%%%%%%%%%%%%%%%%%%%%%%%%%%%%%
%%%%%%%%%%%%%%%%%%%%%%%%%%%%%%%%%%%%%%%%%%%%%%%%%%%%%%%%%%%%%%%
\section*{Acknowledgments}

This work has been partly funded by Regione Lombardia through the initiative ``Programma degli interventi per la ripresa economica: sviluppo di nuovi accordi di collaborazione con le universita' per la ricerca, l'innovazione e il trasferimento tecnologico'' - DGR n. XI/4445/2021.
We recognise the priceless contribution made by several CP, dHCP, ALBERTS, iSeg-19, MCRIB2.

%%%%%%%%%%%%%%%%%%%%%%%%%%%%%%%%%%%%%%%%%%%%%%%%%%%%%%%%%%%%%%%%%%%%%%%%%%%%%%%%%%%%%%%%%%%%%%%%%%%%%%%%%%%%%%%%%%%%%%%%%%%%%%
\bibliographystyle{apalike} 			%unsrt}  
\bibliography{refs.bib}

\end{document}